\documentclass[aps,prb,twocolumn,floatfix,showpacs]{revtex4-2}
\usepackage{amssymb, amsmath, color}
\usepackage[utf8]{inputenc}
\usepackage[dvipsnames]{xcolor}
\usepackage{xcolor}
\definecolor{royal}{HTML}{4169E1}
\usepackage{graphicx}
\usepackage{amsmath,amssymb}
\usepackage{dcolumn} 
\usepackage{hyperref}
\usepackage{ulem}

\hypersetup{%
  colorlinks=true,
  citecolor=blue,
  urlcolor=blue,
  linkcolor=blue,
  }

\begin{document}
\title{Strength of the Hubbard potential and its modification by breathing distortion in $\text{BaBiO}_{3}$}

\author{Alexander E. Lukyanov$^{1}$}
\author{Ivan A. Kovalev$^{1}$}
\author{Vyacheslav D. Neverov$^{1}$}
\author{Yaroslav V. Zhumagulov$^{2,1}$}
\author{Andrey V. Krasavin$^{1}$}
\author{Denis Kochan$^{2}$}

\affiliation{$^{1}$National Research Nuclear University MEPhI, Kashirskoye shosse 31, Moscow, 115409, Russian Federation}

\affiliation{$^{2}$Institute for Theoretical Physics, University of Regensburg, Regensburg, 93040, Germany}

\begin{abstract}
$\text{BaBiO}_{3}$ compound is known as an archetype example of a three-dimensional Holstein model with the realization of the charge density wave state at half-filling and the superconducting state when doped. Although many works are devoted to the study the electron-phonon interaction in $\text{BaBiO}_{3}$, the influence of the electron-electron Hubbard 
interaction on the electronic structure in this system is still under quest. In our work, we obtain analytical expression for the screened Coulomb potential, and along with the 
basis of \textit{ab initio}-computed maximally localized Wannier orbitals, we quantitatively estimate the magnitude of the effective on-site Hubbard potential scrutinizing the 
effects of distortion of the crystal lattice. 
We show that a proper inclusion of the electron-electron interactions into the Holstein model significantly lowers the value of the underlying electron-phonon coupling. Finally, we find 
that the amplitudes of the repulsive electron-electron potential and its attractive counterpart mediated by the electron-phonon coupling are rather comparable. This may open a way for 
a realization of the intermediate phase of $\text{BaBiO}_{3}$ in terms of the Holstein-Hubbard model.
\end{abstract}

\date{\today}
\maketitle

\section{Introduction}
Perovskite compounds are of particular interest because of a large variety of physical phenomena they exhibit. A high degree of chemical functionalization
and structural flexibility in a combination with an inhomogeneous distribution of partially filled d-states lead to a coexistence of several 
interactions that operate on spin, charge, lattice, and orbital degrees of freedom. 
Their mutual interplay generates a wide range of physical properties and functional capabilities counting 
colossal magnetoresistance~\cite{Salamon2001, vonHelmolt1993}, ferroelectricity~\cite{Cohen1992}, superconductivity in cuprates~\cite{Bednorz1986} and bismuthates~\cite{ Sleight1975, Mattheiss1988}, metal--insulator transition~\cite{Imada1998}, ferromagnetism~\cite{Serrate2006}, topological insulators~\cite{Xiao2011}, etc.

\begin{figure}
    \includegraphics[width=0.5\textwidth]{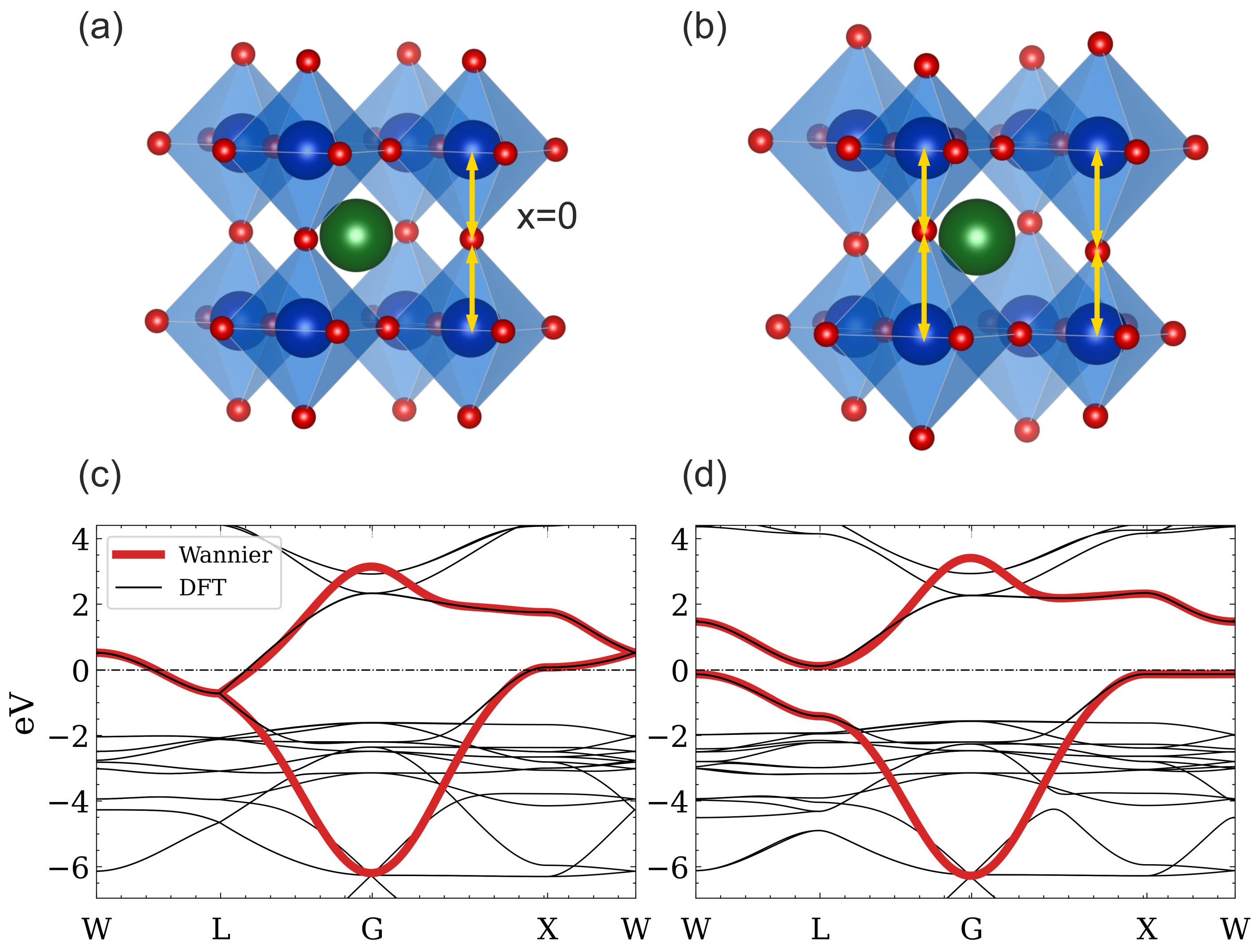}
    \caption{The breathing distortion in $\text{BaBiO}_{3}$ and its effect on the electronic band-structure. $\text{Bi}$ ions are displayed in blue, $\text{O}$ in red, and $\text{Ba}$ in green. Panels (a) and (b) show schematically undistorted and distorted lattice structures with 
    large and small octahedra parameterized by the length of Bi-O bond; yellow arrows indicate changes in their lengths and $x$ stands for the corresponding distortion amplitude. 
    Large octahedra host an additional pair of electrons, while the small ones have a pair of holes; the sizes of octahedra are exaggerated for clarity.  
    Panels (c) and (d) show, correspondingly, the electronic band structures of undistorted and distorted, $x=0.1$\,\AA, lattices; black lines correspond to \textit{\textit{ab initio}} calculations, while the red lines come from the effective tight-binding model (wannierization) that covers two bands near the Fermi level. Direct gap exists for any nonzero 
    distortion, while the indirect one opens for the distortions above $0.1$\,\AA.}
    \label{fig:Fig1}
\end{figure}

Properties of many
$\text{ABX}_{3}$-type perovskites---where A is a large cation, usually an alkaline earth or rare earth element, B is relatively small ion of 3d-, 4d-, or 5d-transition metal, and X are anions, usually oxygen atoms that form the octahedral environment of the B ion, see Fig.~\ref{fig:Fig1}(a)---are extremely 
sensitive to distortion, rotation and tilting of the $\text{BX}_{6}$ octahedra. 
These structural modifications and distortions control hopping amplitudes and exchange interactions through the lengths and angles of the octahedral B--X--B bonds and consequently, the underlying electronic and magnetic properties of the perovskites compounds~\cite{He2010, Chakhalian2011}.

$\text{BaBiO}_{3}$ is a remarkable representative of the perovskite class as it realizes high-temperature superconductivity while does not include transition metal ions~\cite{Mattheiss1988}. In normal conditions, $\text{BaBiO}_{3}$ is characterized by anomalously high amplitude of phonon oscillations leading to pronounced breathing and tilting distortions of the crystal structure~\cite{Uchida1987}. The breathing distortion is caused by a tunneling of electron pairs and subsequent charge transfer between the neighboring octahedra---as a consequence Bi-O bonds change their lengths and the octahedral pattern dynamically alters its proportions along three crystallographic directions~\cite{Sleight1975, Plumb2016, Menushenkov2016}. Recently, it was established that the distortion amplitude in $\text{BaBiO}_{3}$ may be controlled by a strong laser pulse, which drives the system from insulating to metallic state with no distortions~\cite{Lukyanov2020}.

From the theoretical point of view $\text{BaBiO}_{3}$ serves as an archetypal lattice model~\cite{Mattheiss1988, Meregalli1998, Foyevtsova2015} for studying electron-phonon phenomena in perovskites as its electronic properties are defined mostly by a single effective orbital---the molecular Bi-O hybrid. 
The majority of recent works aimed to focus on the Holstein-like models~\cite{Seibold1993,Nourafkan2012, Yam2020, Lukyanov2020}, though some early reports found also a negative 
Hubbard-$U$ parameters by employing first principles calculations~\cite{Vielsack1993} or phenomenological considerations~\cite{Varma1988}.
The use of the Holstein-like models in the above-mentioned studies put forward suggestion that the electron-phonon coupling dominates other interactions in $\text{BaBiO}_{3}$. 
The central aim of this work is to demonstrate that the Coulomb interaction in $\text{BaBiO}_{3}$ is of considerable strength and affects the main electronic properties of the compound 
including the electron-phonon coupling itself, so the Coulomb repulsion should not be neglected. Also, we propose a certain modification of the Coulomb interaction on the model level taking into account an influence of the large distortions of the crystal lattice of $\text{BaBiO}_{3}$ with the electrons' pair tunneling among the neighboring octahedra.

For this purpose, we calculate electronic band structure, perform wannierization and compute matrix elements of the Coulomb interaction in the basis of \textit{ab initio}-computed maximally localized Wannier orbitals---all that for different distortion amplitudes. Based on these calculations we estimate strengths of the underlying Holstein $g$ and Hubbard $U$ parameters that enter the effective lattice model.  
In this work, we restrict ourselves to the breathing mode displayed in Fig.~\ref{fig:Fig1}, since the tilting distortions do not change significantly the electronic properties of $\text{BaBiO}_{3}$~\cite{Khazraie2018}.

The main difficulty associated with calculating the Coulomb matrix elements is of a technical nature: the calculation of the screened Coulomb potential requires exact convergence in many parameters, such as $\textbf{k}$-mesh and the number of conduction bands~\cite{Aryasetiawan2006,Vaugier2012,Nilsson2013,Timrov2018,Kim2021,TancogneDejean2020,Schler2013,Prishchenko2017}. To overcome this obstacle, we use a model of the dielectric function for calculating the screened Coulomb potential. Model dielectric functions were used to describe many-body effects in three-dimensional semiconductors long before the development of computer modeling of materials~\cite{Penn1962, Levine1982, Hybertsen1988, Cappellini1993}; in recent years, this approach has gained relevance in the study of low-dimensional materials, in particular, graphene~\cite{Petersen2009}, graphane~\cite{Cudazzo2011}, two-dimensional heterostructures~\cite{Latini2015}, monolayers of transition metal dichalcogenides~\cite{Trolle2017}, etc. In this work we use the model dielectric function introduced by G.~Cappellini \textit{et al.}~\cite{Cappellini1993} and generalize it to account for non-local effects. It turns out that the integral for the screened Coulomb potential can be calculated analytically in the three-dimensional isotropic case after which calculation of the Hubbard $U$ becomes a relatively simple task.

The paper is organized as follows: section \ref{Model} provides theoretical basis for the methods employed in the sequel of the work, section \ref{Results} presents details of the calculations, and as well, discusses the obtained results, finally, section \ref{Conclusions} summarizes the work and outlines some perspectives.

\section{Model} \label{Model}

\subsection{Model formulation---rationals}

$\text{BaBiO}_{3}$ compound is used widely \cite{Sleight2015,Nourafkan2012,Varma1988,Rice1981} as a textbook example of the Holstein model 
with the Hamiltonian 
\begin{equation}\label{eq:holstein}
    H_H = H_{el}+ H_{ph} + H_{el-ph}
\end{equation}
that takes into account electronic and phononic degrees of freedom and also a strong electron-phonon interaction coupling them. 
The effective model Hamiltonian $H_H$ operates on the single-orbital cubic lattice (at half-filling) formed by Bi sites, since therein-centered Wannier orbitals dominate the $\text{BaBiO}_{3}$ band structure around the Fermi level, see the bold-red bands mainly along W-L and X-W paths in Figs.~\ref{fig:Fig1}(c)~and~(d).
Particular terms of $H_H$ have the following structures:
\begin{align}
    H_{el} &=  -\sum_{ij \sigma}  t_{ij} \left(c^\dagger_{j \sigma}c_{i \sigma}+c^\dagger_{i \sigma}c_{j \sigma} \right),\label{eq:holstein_el}\\
    H_{ph} &=  \omega_{ph}\sum_i b^\dagger_i b_i, \label{eq:holstein_ph}\\
    H_{el-ph} &= g\sum_{i\sigma}\left(b^\dagger_i+b_i\right)n_{i\sigma}. \label{eq:holstein_el_ph}
\end{align}
Here, $c_{i\sigma}^{(\dagger)}$ are the electron operators annihilating (creating) an electron with spin $\sigma$ at site $i$, $t_{ij}$ are the underlying hopping integrals, and $n_{i\sigma} = c_{i\sigma}^\dagger c_{i\sigma}$ are the corresponding number operators. 
Moreover, we consider a single phonon mode with frequency $\omega_{ph}$ described by the phononic annihilation (creation) operators $b_i^{(\dagger)}$ acting on the lattice site $i$. 
This mode represents the optical breathing distortion in $\text{BaBiO}_{3}$ that couples to the electron density via the on-site electron-phonon coupling $g$. Its explicit value in terms of DFT data is given below along with other interaction constants.
Although it is widely acknowledged that the Holstein model describes systems sufficiently well, in our work we would like to go beyond and consider the additional interactions to the Holstein Hamiltonian $H_H$, Eq.~(\ref{eq:holstein}). Particularly, we study the \textit{on-site Hubbard} interaction $H_{U}$ and its associate \textit{on-site Hubbard-phonon} interaction $H_{U-ph}$ triggered by the action of the optical breathing phonon mode: 
\begin{align}
    H_{U}&=U\sum_{i} n_{i\uparrow}n_{i\downarrow} ,\label{eq:on-site-Hubbard}\\
    H_{U-ph}&=\gamma\sum_{i}\left(b^\dagger_i+b_i\right)n_{i\uparrow}n_{i\downarrow}.\label{eq:on-site-Hubbard-phonon}
\end{align}
Consequently, we term $U$ as the on-site Hubbard coupling and $\gamma$ as the on-site Hubbard-phonon coupling. Although the role of the Hubbard term is clear, it is worth to say why we care to introduce the Hubbard-phonon term.
The breathing distortion in $\text{BaBiO}_{3}$ is accompanied by a tunneling of the electron/hole pairs and hence a substantial charge transfer between the neighboring Bi-centered octahedra. For that reason we expect in our ``poor-man extension'' of the Holstein-Hubbard model also a certain coupling between the phonons and the pair-density operator, i.e.~$n_{i\uparrow}n_{i\downarrow}$-term. However, integrating out phonon degrees of freedom, we can down-fold $H_{U-ph}$ and $H_{el-ph}$ Hamiltonians into effective on-site 
electron-electron interactions of the Hubbard-type but with dynamical dependence on $\omega$:
\begin{align}
    H_{U-ph}(\omega)&=\gamma^2\frac{2\omega_{ph}}{\omega^2-\omega_{ph}^2}\sum_{i}n_{i\uparrow}n_{i\downarrow},\\ 
    H_{el-ph}(\omega)&=g^2\frac{2\omega_{ph}}{\omega^2-\omega_{ph}^2}\sum_{i}n_{i\uparrow}n_{i\downarrow}.
\end{align}
It is clear from the above expressions that bringing back phonon degrees of freedom their sum would behave as a sole electron-phonon interaction with the renormalized electron-phonon coupling strength:
\begin{align}
    g^{*}=\sqrt{g^2+\gamma^2}.
\end{align}
Thus, we can conclude that the fluctuations of the on-site Hubbard potential $U$ enhance the electron-phonon coupling constant $g$. It is obvious that the presence of electron-electron interactions have a significant impact on the underlying electronic structure of the system and its phase-diagram---it is known that the Holstein-Hubbard model, $H_H+H_U$, leads to a different phase diagram as that predicted by its less-interactive Holstein ancestor~\cite{Berger1995,Tezuka2007,Johnston2013,Becca1996,Hardikar2007,Clay2005,Assaad2007,Macridin2004,Capone2004,Nowadnick2012,Murakami2013,Werner2007,Costa2020}. In these regards, 
the electron-electron repulsive interaction can contra-act the attractive interaction mediated by electron-phonon coupling, lowering its strength and, possibly, even turning the attraction into an effective repulsion. The aim of our work is to estimate magnitudes of $t_{ij}$, $g$, $U$ and $\gamma$ using \textit{\textit{ab initio}} calculations and analytical results about screening. 

\subsection{Estimation of the model parameters---conceptuals}

The starting point is the \textit{ab-initio} calculation of the electronic band structure of $\text{BaBiO}_{3}$ subjected to different distortion amplitudes of the frozen breathing mode, see Fig.~\ref{fig:Fig1} for the structure visualization and its electronic band structure. We have calculated electronic band structures for eight representative breathing distortions starting from the undistorted system, for more details see section \ref{Results}. 
\begin{figure}
    \includegraphics[width=0.48\textwidth]{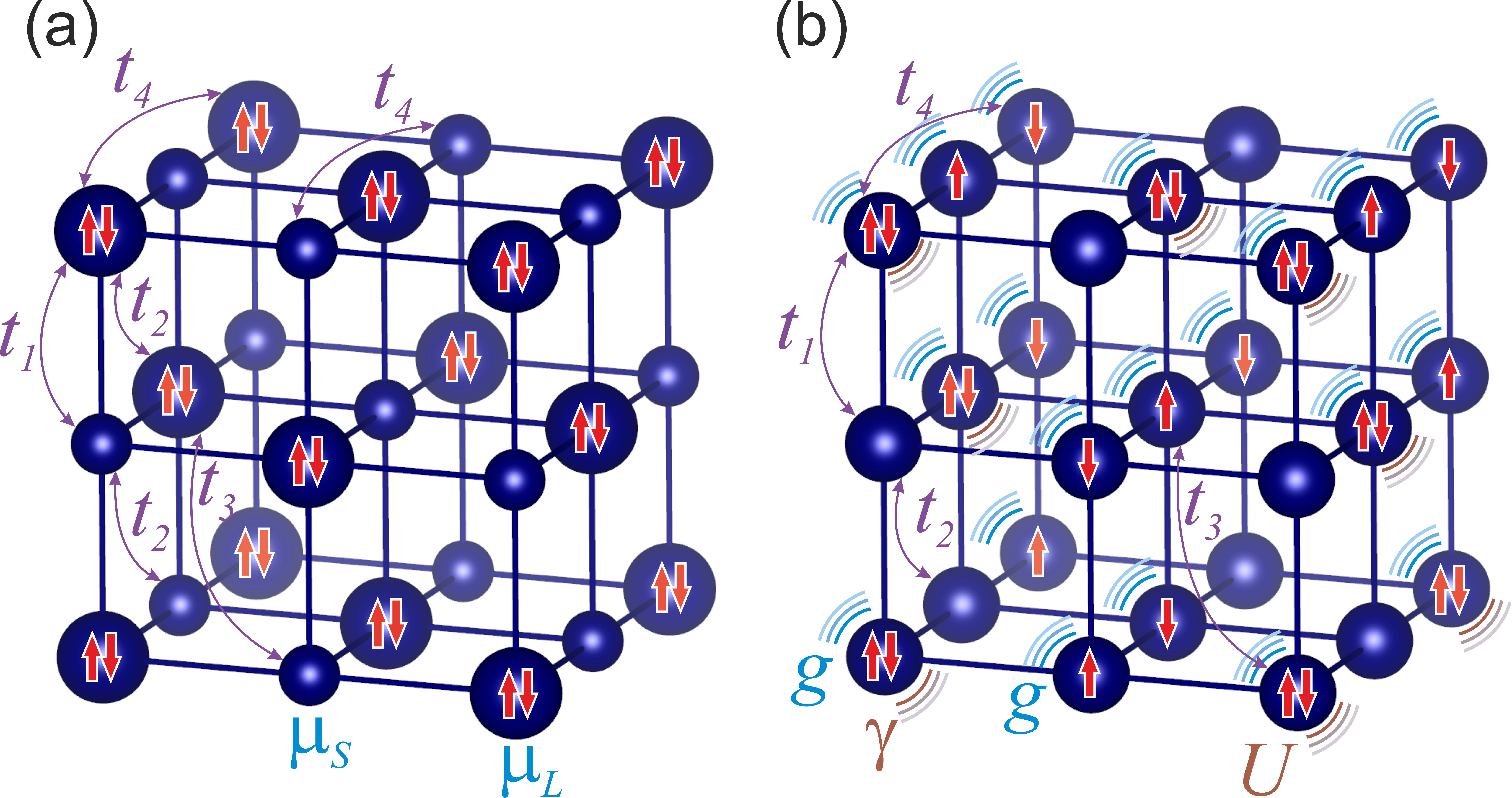}
    \caption{Schematic mapping of the single-orbital tight-binding Hamiltonian onto the interacting Holstein--Hubbard model. (a) The effective single-orbital tight-binding model Hamiltonian is defined on a cubic lattice with the non-equivalent s and L nearest-neighbor Bi-sites with different on-site energies $\mu_{\textrm{s}}$ and $\mu_{\textrm{L}}$. 
    The hopping integrals $t_i$ are taken into account up to the forth coordination number. In reality this picture dynamically evolves in time and hence s- and L-octahedra occupy 
    different lattice sites.
    To visualize a fact that the large octahedra host extra electron pairs the L-sites (with on-site energy $\mu_{\textrm{L}}$) are displayed as pair occupied. 
    (b) The interacting Holstein-Hubbard model is defined on a cubic lattice with equivalent sites but varying electron occupancies. Singly occupied sites give rise to the 
    Holstein electron-phonon interaction with the coupling strength $g$, while the doubly occupied sites interact also via the Hubbard and the Hubbard-phonon interactions parameterized by $U$ and $\gamma$. Kinetic part of the Holstein-Hubbard model involving single-electron hoppings $t_i$ is the same as in panel (a), phonons
    are displayed as concentric blue-brown wave-echos around Bi sites.}
    \label{fig:Mod}
\end{figure}

Since we aim to capture physics around the Fermi level we focus on two closest bands in its vicinity---the central bands evolving mainly along W-L and X-W paths in Figs.~\ref{fig:Fig1}(c)~and~(d) spanning roughly the energy window from $-2$~to~$2$~eV. These two bands are 
further subjected to the wannerization procedure that gives us an effective single-orbital tight-binding 
model Hamiltonian $H_{TB}$ acting on the cubic lattice with two non-equivalent nearest-neighbor Bi-sites, see Fig.~\ref{fig:Mod}. 
This is quite reasonable since the two neighboring Bi atoms within $\text{BaBiO}_{3}$ are surrounded by two spatially different octahedra, and hence different charge environments. We call the Bi-sites surrounded by the small (compressed) and large (expanded) octahedra as the small and large sites, reserving for them the subscripts ``s'' and ''L,'' correspondingly, when necessary, we talk also about the ``s'' and ``L'' sublattices of the cubic crystal. 
The effective tight-binding Hamiltonian $H_{TB}$ stemming from the wannierization is parameterized by the hopping integrals $t_{ij}$ and the on-site energies $\mu_{\mathrm{s}}$ and $\mu_{\mathrm{L}}$,
\begin{align}
    H_{TB} =  &-\sum_{ij \sigma}  t_{ij} \left(c^\dagger_{j \sigma}c_{i \sigma}+c^\dagger_{i \sigma}c_{j \sigma} \right)\nonumber\\
    &+\mu_{\mathrm{s}}\sum_{i\in\mathrm{s},\sigma} c^\dagger_{i \sigma}c_{i \sigma}
    +\mu_{\mathrm{L}}\sum_{i\in\mathrm{L},\sigma} c^\dagger_{i \sigma}c_{i \sigma}, \label{Eq:TB-Ham}
\end{align}
whose values are summarized in Table~\ref{tab:parameters} for different distortion strengths $x$. The sum in the first (kinetic) term of $H_{TB}$ runs over the first up to the forth-nearest neighbors, and the corresponding hopping integrals among them are denoted as $t_1,\dots,t_4$ in Table~\ref{tab:parameters}. 
As a comment, due to the non-equivalent ``s'' and ``L'' sites, the cubic unit cell of the tight-binding 
model is twice as large as the corresponding cubic unit cell of the Holstein model. For the latter the breaking of the sublattice symmetry appears dynamically as a result of the electron-phonon coupling with the breathing optical phonon mode. 

Since the hoppings integrals entering $H_{TB}$, Eq.~(\ref{Eq:TB-Ham}), are relatively weakly dependent on the distortion 
$x$, we can naturally identify the kinetic part of the Holstein model, Hamiltonian $H_{el}$, Eq.~(\ref{eq:holstein_el}), with the kinetic part of the wannerization-generated tight-binding model, i.e.
\begin{equation}
    H_{el}=\text{kinetic term of }H_{TB}.
\end{equation}
Contrary, the on-site energies $\mu_{\mathrm{s}}$ and $\mu_{\mathrm{L}}$ are very strongly distortion dependent, meaning the electron-phonon coupling with the breathing mode in $\text{BaBiO}_{3}$ is sufficient. On the level of Holstein model
such coupling is governed by the Hamiltonian $H_{el-ph}$, Eq.~(\ref{eq:holstein_el_ph}), where the coupling constant $g$
can be estimated as follows:
\begin{align}
g=\frac{\partial \mu}{\partial x}  \,\sqrt{\frac{1 }{2M_O\omega_{ph}}},
\label{eq:g}
\end{align} 
where $\mu=(\mu_{\mathrm{s}}-\mu_{\mathrm{L}})/2$ is the (interpolated) difference (function) of two on-site energies entering $H_{TB}$, $M_O$ is the oxygen mass, $ \omega_{ph}=70$~meV~\cite{Sugai1985, Tajima1992} is the optical phonon energy and $x$ is the amplitude of the breathing distortion. 

The second outcome of the wannierization are maximally-localized Wannier orbitals $w_{i,\mathrm{s}}$ and $w_{i,\mathrm{L}}$ centered on
different ``s'' and ``L'' lattice sites. In what follows we pickup two neighboring sites and use them-centered
Wannier orbitals---we denote them as $w_{\mathrm{s}}$ and $w_{\mathrm{L}}$---for the calculation of the Hubbard and Hubbard-phonon coupling strengths. 
For the iso-surface visualization of $w_{\mathrm{s}}$ and $w_{\mathrm{L}}$ for the undistorted and distorted lattice see Figs.~\ref{fig:wann}(a)~and~(b).

Calculating the matrix elements of the Coulomb interaction in terms of the tight-binding model stemming from the \textit{ab initio} calculation is a computationally complex problem, since it requires calculation of the screened Coulomb potential by the random phase approximation method \cite{Hybertsen1986,Hybertsen1988,Aryasetiawan2006,Nilsson2013}.
To get rid over this problem, we use the model dielectric function from \cite{Cappellini1993},
\begin{equation}
\label{eq:dielectric}
    \epsilon(\mathbf{q},\rho(\mathbf{r})) = 1 + \left[ \frac{1}{\epsilon_{\infty}-1}+ \alpha \frac{q^2}{q_{TF}^2} +\frac{ ^2q^4}{4m^2\omega_{p}^2} \right]^{-1},
\end{equation}
where $\epsilon_{\infty}$ is the long-wave length ($q=0$) limit of the dielectric constant specific for each material, ${q_{TF}}$ and $\omega_{p}$ are, correspondingly, the Thomas-Fermi wave vector and the plasma frequency that are functions of the local electron density $\rho(\textbf{r})$ and, finally, $\alpha$ is a dimensionless fitting parameter specific for each material. In our calculations, following Ref.~\cite{Cappellini1993}, we use $\alpha = 1.563$. 
The above choice of the dielectric function is because of its relative simplicity, analytical possibility of deriving the screened potential explicitly, and physically appropriate description of the screening on both long- and short-ranges.

The matrix elements of the screened Coulomb interaction between two Wannier orbitals $w_n$ and $w_m$ ($n,m$ run through $\{\mathrm{s,L}\}$) are expressed as
\begin{align} 
\label{eq:U_nm}
W_{nm}&=\frac{1}{2}\int d\textbf{r} d\textbf{r}' |w_n(\textbf{r})|^2 W_{}\left(\textbf{r}-\textbf{r}',\rho(\textbf{r})\right) |w_m(\textbf{r}')|^2\\  \nonumber
&+\frac{1}{2}\int d\textbf{r} d\textbf{r}' |w_n(\textbf{r})|^2 W_{}\left(\textbf{r}-\textbf{r}',\rho(\textbf{r}')\right) |w_m(\textbf{r}')|^2 ,
\end{align}
where $W=\epsilon^{-1}\,V$ is the screened Coulomb potential by means of the above dielectric function. 
Within this approach, the matrix elements $W_{nm}$ can be calculated without significant computational costs. This is because the difference between the screened, $W$, and the bare, $V$, Coulomb interaction, the so-called \textit{correlated Coulomb potential}, $W_c = W - V$, can be represented in the explicitly analytical form. Performing Fourier transformation of $W_c$~\cite{Hybertsen1986},
\begin{equation} \label{V_corr}
   W_c(\textbf{r}) = \frac{1}{(2\pi)^3}\int d^{3}\mathbf{q}\frac{4\pi e^2}{q^2}\Bigl[\frac{1}{\epsilon(\mathbf{q},\rho(\mathbf{r}))}-1\Bigr]e^{i\mathbf{q}\mathbf{r}},
 \end{equation} 
one arrives in the case of spherical symmetry at final expression that counts just material parameters 
and the magnitude of spatial distance $r=|\textbf{r}|$:

\begin{align} 
    W_c(r) &=- \frac{e^2}{2r}\Bigl(1 - \frac{1}{\epsilon_{\infty}} \Bigr) \times \\ \nonumber & \mathrm{Re} \left\{ 2 - \frac{1+A}{A} e^{-\kappa r\sqrt{1-A}} + \frac{1-A}{A} e^{-\kappa r \sqrt{1+A}} \right\},
\end{align}

or
\begin{widetext} 
    \begin{equation}
    W_c(r)=\begin{cases}
    -\frac{e^2}{2r}\left(1 - \frac{1}{\epsilon_{\infty}} \right)\left[2 - \frac{1+A}{A} e^{-\kappa r\sqrt{1-A}} + \frac{1-A}{A} e^{-\kappa r \sqrt{1+A}} \right]; & \zeta<1,\\
    -\frac{e^2}{r}\left(1 - \frac{1}{\epsilon_{\infty}} \right) \left[ 1-e^{-\kappa ra} \left( \cos{\kappa rb} +\frac{1}{A} \sin{\kappa rb}\right) \right]; & \zeta > 1,
    \ a = \sqrt{\frac{\sqrt{\zeta} + 1}{2}},
    \ b=\sqrt{\frac{\sqrt{\zeta} - 1}{2}}
    \end{cases}
    \label{eq:W_c}
    \end{equation}
\end{widetext}
with
\begin{equation}
     A=\sqrt{|1-\zeta|},\ 
    \kappa=\frac{(2\alpha)^{1/2}\pi^{2/3}}{3^{1/6}}\,\rho(\textbf{r})^{1/3}. 
\end{equation}
Here $\zeta$ is the dimensionless parameter,
\begin{equation}
    \zeta =\frac{\xi}{a_0 \rho(\textbf{r})^{1/3}},
\end{equation}
where $a_0$ is the Bohr radius and 
\begin{equation}
    \xi = \frac{4 \cdot 3^{2/3}}{\pi^{5/3} \alpha^2} \frac{\epsilon_{\infty}}{\epsilon_{\infty}-1}.
\end{equation}
The detailed derivation of Eq.~(\ref{eq:W_c}) is presented in the Appendix.

The matrix elements $W_{nm}$ are then obtained as sum of the matrix elements of the correlated Coulomb potential, $W_{c,nm}$, and those of the bare Coulomb interaction,
\begin{align} 
\label{eq:V_nm}
V_{nm}&=\int d\textbf{r} d\textbf{r}' |w_n(\textbf{r})|^2 V_{}\left(\textbf{r}-\textbf{r}'\right) |w_m(\textbf{r}')|^2. 
\end{align}
As a comment, $W_{nm}$ depends on the distortion $x$, since the later affects ``shapes'' of the Wannier orbitals $w_n$ and $w_m$, see Fig.~\ref{fig:wann}, and also the charge distribution $\rho(\mathbf{r})$ entering the dielectric function $\epsilon$, Eq.~(\ref{eq:dielectric}).

In the next step, the self-screening issues should be properly taken into account. An electron cannot screen itself, which makes calculations using the random phase approximation inappropriate. Therefore, it is conceptually erroneous to estimate the Hubbard $U$ 
parameter in terms of the fully screened Coulomb potential $W$. To avoid the double screening, one usually uses the constrained random phase approximation (cRPA) \cite{Aryasetiawan2006}, which is a sort of the unscreening procedure that takes fully screened Coulomb matrix elements $W_{nm}$ 
and provides their partly-unscreened-representatives, $W^r_{nm}$, facilitated for the bands defined by the effective tight-binding model. 
Within cRPA, the matrix elements of the Coulomb interaction, $W^r_{nm}$, can be expressed in the following way \cite{Aryasetiawan2006,Schler2013,Wehling2011,Vaugier2012,Nilsson2013}:
\begin{align}\label{eq:W^r}
    W^r=W\left[I+P^{d}\,W\right]^{-1},
\end{align}
where $W$ represents the fully screened Coulomb interaction, Eq.~(\ref{eq:U_nm}), and $P^{d}$ is the band subspace polarization specified by entries of the effective tight-binding model. Down-to-earth, the band subspace polarization can be calculated as \cite{Levine1982,Hybertsen1986,Aryasetiawan2006,Prishchenko2017}
\begin{align}
    P^{d}_{nm}&=\frac{g_s}{N^2_k}\sum_{\textbf{kq},\alpha \beta} \frac{f_\beta(\textbf{k})-f_\alpha(\textbf{k}+\textbf{q})}{E_\beta(\textbf{k})-E_\alpha(\textbf{k}+\textbf{q})}\\ \nonumber
    &\times C_{n,\beta}(\textbf{k})C^{*}_{n,\alpha}(\textbf{k}+\textbf{q})
    C^{*}_{m,\beta}(\textbf{k})C_{m,\alpha}(\textbf{k}+\textbf{q}),
\end{align}
where $g_s=2$ is the spin degeneracy factor, $N_k$ is the number of \textbf{k}-points in the Brillouine zone, $n,m\in\{\mathrm{s, L}\}$, sum over $\alpha$ and $\beta$ runs over the valence and conduction bands of the underlying tight-binding model, $E_\alpha(\textbf{k})$ is the tight-binding single-particle energy, $f_\alpha(\textbf{k})$ is the corresponding Fermi--Dirac occupation, and  $C_{n,\alpha}(\textbf{k})$ is the probability amplitude giving contribution of the $n$th Wannier orbital to the $\alpha$-th Bloch-band-state with momentum $\textbf{k}$.
Finally, equipped with the partly-screened Coulomb matrix elements $W^r_{nm}$, we calculate~\cite{Wehling2011,Schler2013} the optimal values of the on-site Hubbard $U$ parameter and the Hubbard-phonon coupling $\gamma$: 
 \begin{align}
     U_\mathrm{s}&=W^r_{\mathrm{ss}}-W^r_{\mathrm{sL}},\ \ \ U_\mathrm{L}=W^r_{\mathrm{LL}}-W^r_{\mathrm{sL}}, \label{eq:UsL}\\
     U&=\frac{1}{2}\left<U_\mathrm{s}+U_\mathrm{L}\right>_{x}, \label{eq:U}\\
     \gamma&=\frac{\partial U}{\partial x}  \,\sqrt{\frac{1 }{2M_O\omega_{ph}}}\label{eq:gamma}.
 \end{align}
As a comment, $U_{\mathrm{s/L}}$ and $U$ are understood as the interpolated functions of the distortion $x$ that is entering $W^r_{nm}$ and from them derived quantities. Symbol $\left<A\right>_x$ represents an average value of the quantity $A$ taken over the range of crystal lattice distortions.

In this work, we consider a single-orbital tight-binding model, since it fully incorporates physics of the $\text{BaBiO}_{3}$ near the Fermi energy level. However, all of the above equations and concepts can be naturally extended to involve multi-orbital models.

\begin{center}
\begin{table}
    \centering
\begin{tabular}{|c|c|c|c|c|c|c|}
 \hline
  $x$ [\AA] & $\mu_\mathrm{s}$ [eV] & $\mu_\mathrm{L}$ [eV] & $t_1$ [eV] & $t_2$ [eV] & $t_3$ [eV] & $t_4$ [eV] \\
 \hline
 \hline
     0.000 & -0.088 & -0.088 & -0.514 & -0.130 & -0.056 & 0.078 \\
     \hline
     0.100 & 0.758 & -0.808 & -0.542 & -0.151 & -0.049 & 0.085 \\
     \hline
     0.125 & 0.972 & -1.000 & -0.539 & -0.170 & -0.063 & 0.083 \\
     \hline
     0.150 & 1.199 & -1.125 & -0.532 & -0.177 & -0.061 & 0.088 \\
     \hline
     0.175 & 1.401 & -1.256 & -0.524 & -0.187 & -0.062 & 0.097 \\
 \hline
\end{tabular}
    \caption{Effective model parameters for $\text{BaBiO}_{3}$ compound subjected to various amplitudes of the breathing distortion parameterized by $x$. 
    Tight-binding model stemming from the wannierization, Eq.~(\ref{Eq:TB-Ham}), involves the on-site energies
    $\mu_\mathrm{s}$ and $\mu_\mathrm{L}$ at two non-equivalent Bi sites, and the hoppings integrals $t_i$, $i=1,...,4$, ranging among the corresponding 
    $i$-th nearest-neighbors within the underlying cubic lattice, for a schematic see Fig.~\ref{fig:Mod}.
    On-site energies $\mu$'s scale linearly with $x$, while $t_i$'s are only mildly fluctuating at second decimal place. To keep table concise we only show data for a limited set of $x$, the values for remaining $\mu$'s can be easily interpolated.} 
    \label{tab:parameters}
\end{table}
\end{center}

\begin{figure}
    \includegraphics[width=0.4\textwidth]{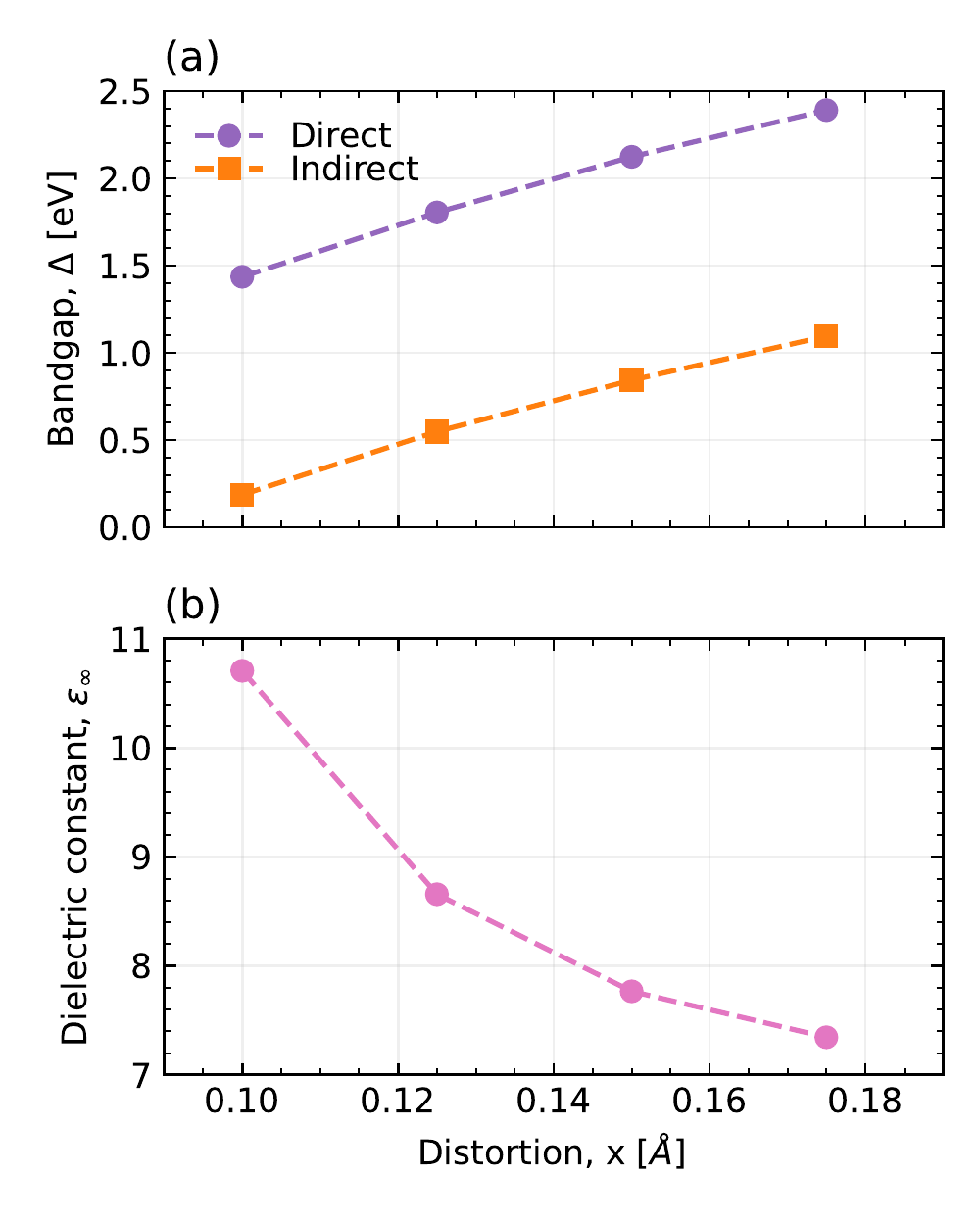}
    \caption{Dependence of the direct and indirect bandgaps, panel~(a), and long-range dielectric constant $\epsilon_{\infty}$, panel~(b), on the $\text{BaBiO}_{3}$ breathing distortion amplitude $x$. 
    The direct gap (circles) opens gradually with the appearance of the distortion, while the indirect gap (squares) is absent and emerges only for $x$ above 0.1~\AA. Correspondingly,
    in the metallic state the system has an infinite dielectric constant $\epsilon_{\infty}$, which is then descending towards finite values---larger the indirect gap, the more insulating system behaves and hence the smaller is $\epsilon_{\infty}$. 
    }
    \label{fig:dft}
\end{figure}

\begin{figure}[h]
    \includegraphics[width=0.3\textwidth]{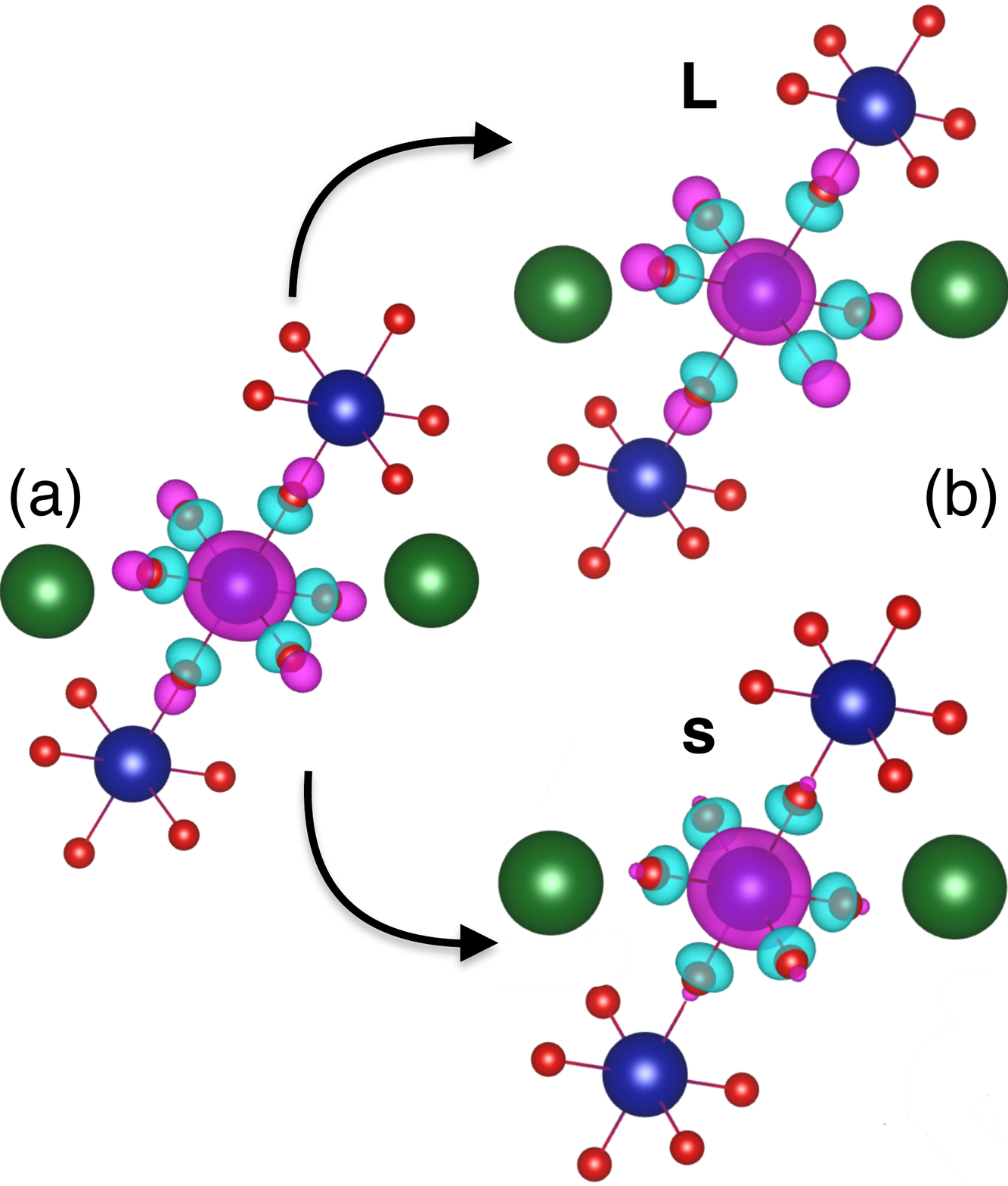}
    \caption{Iso-surfaces of the underlying Wannier orbitals $w_{\mathrm{s}}$ and $w_{\mathrm{L}}$. 
    Panels~(a)~and~(b) show, correspondingly, the iso-surfaces of the undistoreted, $x=0$, and distorted, 
    $x= 0.175$~\AA, Wannier orbitals. In panel~(a) both iso-surfaces coincide, while in panel (b) the upper 
    iso-surface corresponds to $w_{\mathrm{L}}$ and the lower for $w_{\mathrm{s}}$, all the iso-surfaces 
    are displayed for the same iso-level.
    Positive and negative values are shown in light blue and pink, respectively; the colors of atoms are the same as in Fig.~\ref{fig:Fig1}. }
    
    \label{fig:wann}
\end{figure}

\begin{figure}[h]
    \centering
    \includegraphics[width=0.4\textwidth]{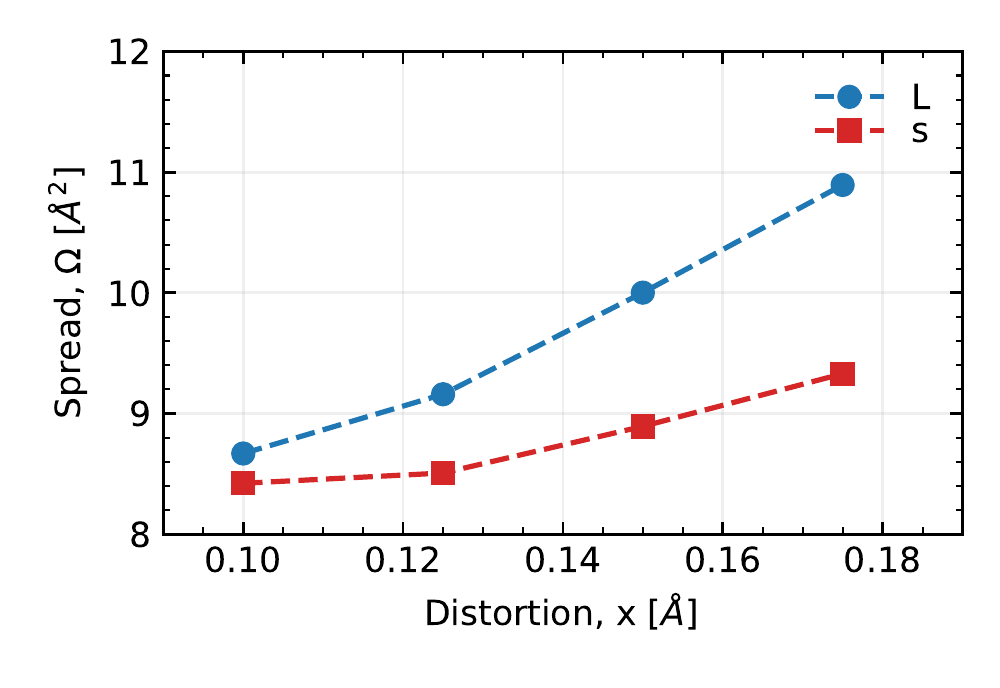}
    \caption{Spreadings, $\Omega=\langle \mathbf{r}^2\rangle-\langle \mathbf{r}\rangle^2$, of the Wannier functions depending on the distortion. Red and blue colors correspond to $w_{\mathrm{s}}$ and $w_{\mathrm{L}}$ Wannier states; dashed lines are guides to the eye.}
    \label{fig:spread}
\end{figure}

\begin{figure}[h]
    \includegraphics[width=0.4\textwidth]{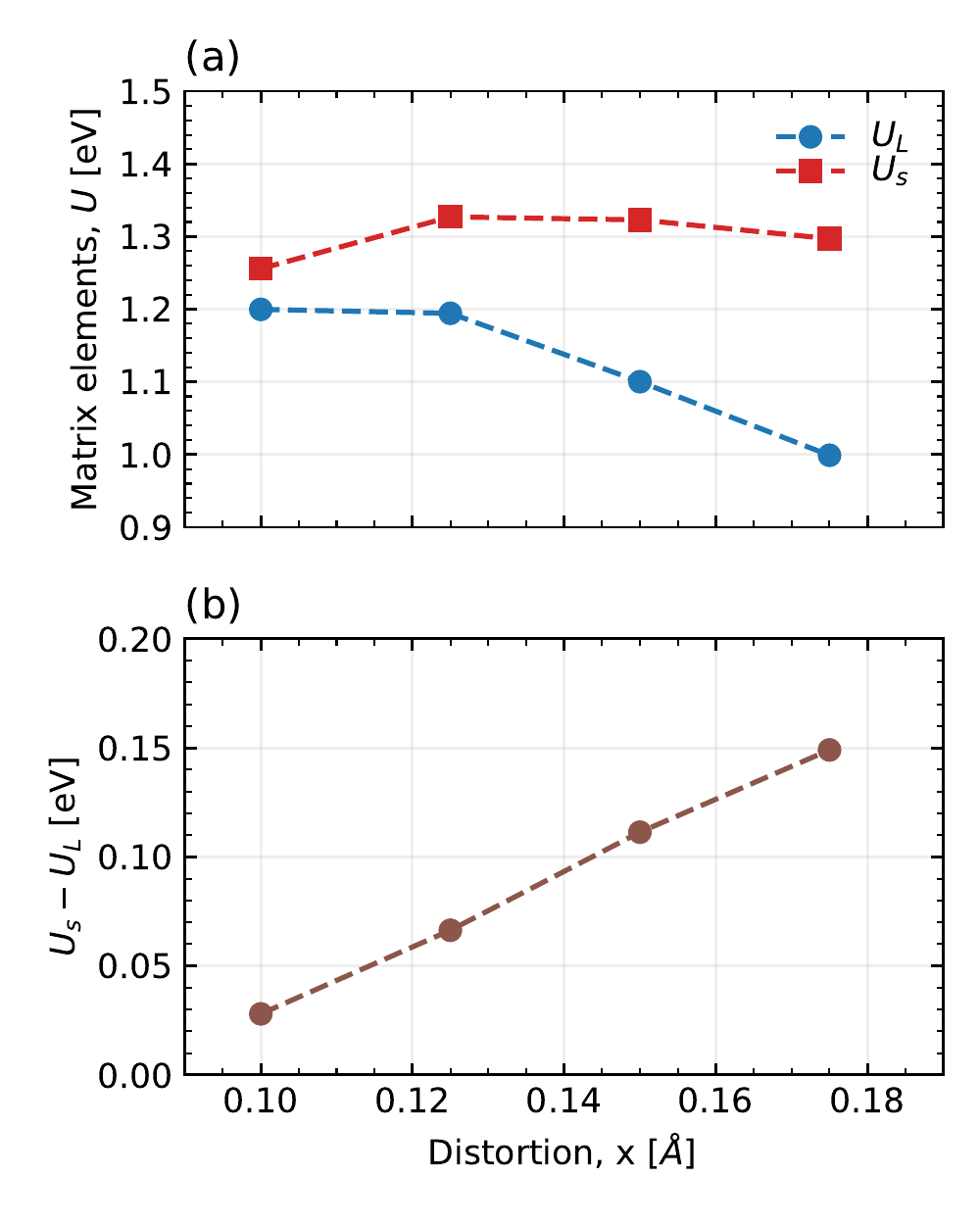}
    \caption{Evolution of the Hubbard parameters $U_{\mathrm{s}}$ and $U_{\mathrm{L}}$, panel (a), and their difference $U_{\mathrm{s}}-U_{\mathrm{L}}$, panel (b), with distortion.}
    
    \label{fig:coulomb}
\end{figure}

\section{Results and computational details} \label{Results}

The basis for estimating the model parameters entering the Holstein, Eq.~(\ref{eq:holstein}), and the Hubbard Hamiltonian, Eq.~(\ref{eq:on-site-Hubbard}), is built 
on the Wannier orbitals that were obtained from the \textit{ab initio} calculations using numerical packages \textit{Quantum Espresso} \cite{Giannozzi2009} and \textit{Wannier90} \cite{Pizzi2020}. 
For bismuth, barium, and oxygen atoms, the norm-conserving semi-relativistic LDA pseudopotentials from the Pseudo Dojo database~\cite{Dojo} were used and the energy cutoff for the plane-wave basis was set at 120~Ry with convergence accuracy of $10^{-8}$. Frozen lattice distortion due to the breathing mode leads to the doubling (``s'' and ``L'' sites) of the unit cell along the (111) direction owing the underlying crystal the cubic face-centered lattice. We have performed eight calculations for the eight representative values of the breathing distortion $x$---starting from the undistorted metallic system and continuing by varying $x$ in steps of 0.025~\AA. 
For distortions below $0.1$~\AA~the system is metallic, see Figs.~\ref{fig:Fig1}(c)~and~(d), and possesses a non-zero direct-gap, however, going 
above~$0.1$~\AA~it turns into an indirect-gap insulator---for the evolution of the direct and indirect gaps with the distortion see Fig.~\ref{fig:dft}(a). Our findings are in 
good agreement with the known experimental result \cite{Uchida1987}  affirming $\text{BaBiO}_{3}$ compound to be the indirect-gap insulator for $x = 0.1$ \AA.

To proceed further, the two bands in the vicinity of the Fermi level, see Figs.~\ref{fig:Fig1}(c)~and~(d), were subjected to the wannerization procedure that provided us 
an effective single-orbital (per Bi atom) tight-binding Hamiltonian $H_{TB}$, Eq.~(\ref{Eq:TB-Ham}), reliable in the energy window from $-2$ to $2$~eV. 
The obtained electron hopping integrals $t_{1},\dots,t_4$, Table \ref{tab:parameters}, and the derived value of the electron-phonon coupling constant, Eq.~(\ref{eq:g}), 
\begin{equation}
    g=0.32\pm0.01\ \text{[eV]}, 
\end{equation}
agree well with the known results~\cite{Khazraie2018}.
Wannierization yielded us also Wannier orbitals $w_\mathrm{s}$ and $w_{\mathrm{L}}$, whose iso-surface plots are displayed for the undistorted and distorted cases in Fig.~\ref{fig:wann}. 
As it is seen, the Wannier orbitals are centered on the Bi atoms and form molecular-hybrids counting $s$-orbital from the bismuth and $p$-orbitals from the surrounding oxygens (directed towards the Bi atoms). The spread, $\Omega$, of the Wannier orbitals---defined as the expectation value $\Omega=\langle \mathbf{r}^2\rangle -\langle \mathbf{r}\rangle^2$---increases with the increased lattice distortion, but not identically for both Wannier states. 
As it is clear from Fig~\ref{fig:spread}, the spread of $w_\mathrm{L}$ increases more significantly than that of $w_{\mathrm{s}}$, what affects a scaling of the underlying Coulomb interaction with $x$.

Our ultimate goal is the calculation the Hubbard parameters $U_{\mathrm{s}}$ and $U_{\mathrm{L}}$, for which we need apart of the Wannier states discussed above also the long-range dielectric constant $\epsilon_{\infty}$, see Eqs.~(\ref{eq:dielectric})~and~(\ref{eq:W_c}). The values of $\epsilon_{\infty}$ for the gapped systems, i.e.~for $x>0.1$~\AA, are obtained by the DFPT method built-in the numerical package \textit{Quantum Espresso} \cite{Giannozzi2009}. 
The results are shown in Fig.~\ref{fig:dft}; with an increased amplitude of the lattice distortion the value of $\epsilon_{\infty}$ decreases---blue triangle-up symbols---while
the size of the indirect-gap rises---red dot symbols---obviously, as the system becomes more insulating the dielectric screening softens.
In the metallic state, $x<0.1$~\AA, the dielectric screening is strong (infinite). 

Equipped with the dielectric constant $\epsilon_{\infty}$ and the Wannier orbitals $w_{\mathrm{s}}$ and $w_{\mathrm{L}}$ we are ready to calculate the bare and screened Coulomb matrix elements $V_{nm}$ and $W^r_{nm}$ for the further estimation of the Hubbard parameters $U_{\mathrm{s}}$ and $U_{\mathrm{L}}$. Increased spread of the Wannier orbitals, Fig.~\ref{fig:spread}, lowers magnitudes of the Hubbard parameters and also counterbalances the screening in the system by decreasing the dielectric constant. Summing up these two effects, we obtain a slightly nonlinear dependence of the Hubbard parameters on the amplitude of the crystal lattice distortion, Fig.~\ref{fig:coulomb}(a). However, the difference between local Hubbard parameters $U_{\mathrm{s}}-U_{\mathrm{L}}$ is strictly linear, Fig.~\ref{fig:coulomb}(b), which is consistent with the above Hamiltonian $H_{U-ph}$ Eq.~(\ref{eq:on-site-Hubbard-phonon}).

The values of the on-site Hubbard potential $U$ and the on-site Hubbard-phonon coupling $\gamma$ that are naturally  
emerging from the non-local screening and the coupling to the breathing phonon mode, can be estimated based on Eqs.~(\ref{eq:U})~and~(\ref{eq:gamma}). Using obtained values of $W^r_{nm}(x)$ and the reference energy of the breathing Raman mode, $ \omega_{ph}=70$\,meV~\cite{Sugai1985, Tajima1992}, the resulting values read
\begin{align}
 U&=\left<\tfrac{1}{2}[W^r_{\mathrm{ss}}+W^r_{\mathrm{LL}}]-W^r_{\mathrm{sL}}\right>_{x}=1.21 \pm 0.12~\text{[eV]};\\
\gamma&=\left(1.63\pm0.04\right)~\text{[eV/\AA]}\times 0.043~\text{[\AA]}\nonumber\\ 
      &=0.070\pm0.002~\text{[eV]}.
\end{align}
To estimate the error, we used the three sigma rule. Thus, we can calculate the correction to the electron-phonon coupling constant as
\begin{align}
    g^{*}=\sqrt{g^2+\gamma^2}=0.33\pm0.01 ~\text{[eV]}.
\end{align}
It is well known that the Hubbard and Holstein interactions---parameterized by $U$ and $g$ couplings---act on an electronic system at the half filling in the opposite way. The Hubbard interaction tends to transfer the system into the Mott insulator regime with a uniform charge distribution, while the Holstein interaction strives to form a charge density wave (CDW) 
with a spatially non-uniform charge rearrangement.
In the effect, the repulsive Hubbard interaction gets contra-acted by an attractive electron-electron interaction resulting from the Holstein coupling of electrons and phonons. The final on-site potential---in the anti-adiabatic approximation---can be quantified by the following expression \cite{Costa2020,Berger1995}:
\begin{align}\label{eq:U*}
  U^{*}:= U-\frac{2g^{*2}}{ \omega_{ph}}=U - \lambda D=:-\lambda^{*}D,
\end{align}
where $\lambda=2g^{*2}/({\omega_{ph} D})$ is the dimensionless strength of the Holstein~$+$~Hubbard-phonon attraction and $D$ is the effective bandwidth, for our model $D=4$\,eV. 
The resulting on-site electron-electron interaction $U^{*}$ can be repulsive or attractive giving rise to the Mott insulator or CDW physics. In the latter case, the right-hand side of 
Eq.~(\ref{eq:U*}) defines an effective dimensionless ``electron-phonon strength'' 
$\lambda^{*}=\lambda+\delta\lambda=\lambda-U/D<\lambda$, whose value gets reduced by $\delta\lambda= -U/D$. 
The final estimations give
\begin{align}
  U^{*} &= U-\frac{2g^{*2}}{ \omega_{ph}}=\left(1.21-3.06\right)\,\text{[eV]}=-1.85\,\text{[eV]}\\
  \lambda^{*}&= \lambda+\delta\lambda=0.765-0.3025=0.4625.
\end{align}
So, we conclude that taking into account the Hubbard repulsion $U$, the effective on-site electron-electron interaction $U^{*}$ is significantly suppressed as compared to the  
Holstein and Hubbard-phonon attractions $\lambda D=2g^{*2}/({ \omega_{ph}})$. Moreover, comparing $\lambda^{*}$ with $\lambda$ we see that such suppression is about 40 percent.

These results show that by neglecting the Coulomb interaction the value of $\lambda$ would be overestimated, and hence also the value of the bandgap of 
$\text{BaBiO}_{3}$ realizing CDW phase. Matrix elements of the Hubbard potential can show their significance for the description of physics of the electronic structure of 
$\text{BaBiO}_{3}$ and for analyzes of the phase transitions in the Holstein-Hubbard model \cite{Becca1996,Hardikar2007,Clay2005,Capone2004,Murakami2013,Costa2020,Tezuka2007} separating the anti-ferromagnetic ordering of the Mott insulator from that of CDW phase. Perhaps, the interplay of two interactions, an attractive electron-phonon and a repulsive electron-electron with close to equal strength may lead to the realization of the high-temperature superconductivity of the $\text{BaBiO}_{3}$ compound.

\section{Conclusions and Perspectives} \label{Conclusions}

In our work, we estimated the magnitude of the on-site Hubbard potential in the $\text{BaBiO}_{3}$ using the analytical form of the screened Coulomb potential and the basis of the \textit{ab initio}-computed maximally localized Wannier orbitals. Our results show that the magnitude of the repulsive electron-electron Hubbard potential is comparable to the attractive potential of the electron-phonon~$+$~Hubbard-phonon interactions. Therefore, their effective dimensionless attractive constant $\lambda^*$ gets lowered by 40 percent. This indicates a possible significant contribution of the electron-electron interaction to the electronic structure of $\text{BaBiO}_{3}$. 
Although this modification of the Hubbard potential is difficult to take into account in the framework of the standard Migdal--Eliashberg theory~\cite{Giustino2017}, we assume a possible implementation of quantum Monte Carlo methods~\cite{Nowadnick2012,Johnston2013}, taking into account these corrections. The electron-phonon interaction and the electron-electron interaction have the same order of magnitude in our calculations, which indicates the possibility of realization of an intermediate phase between the anti-ferromagnetic ordering of the Mott insulator and the band insulator phase of the charge density wave. This opens the way for further research in the field of the origin of the superconducting state in $\text{BaBiO}_{3}$ compound.

\section*{Acknowledgements}
The work was supported by the Russian Foundation for
Basic Research under the Project 18-02-40001 mega and by the German Science Foundation---Project DFG~SFB~1277. 
Y.\,V.\,Z. is grateful to the Deutsche Forschungsgemeinschaft (DFG, German Research Foundation) SPP 2244 (Project-ID 443416183) for the financial support. 
The calculations were performed using resources of NRNU MEPhI high-performance computing center.

\section*{Appendix: Derivation of $W_c$}
Below we describe the main steps in evaluating the following integral:
\begin{equation} \label{V_corr_sup}
    W_c(\textbf{r}) = \frac{1}{(2\pi)^3}\int d^{3}\mathbf{q}\frac{4\pi e^2}{q^2}\Bigl[\frac{1}{\epsilon(\mathbf{q},\rho(\mathbf{r}))}-1\Bigr]e^{i\mathbf{q}\mathbf{r}}. 
\end{equation}
Since the integral is spherically symmetric it can be expressed in the following way:
\begin{align}
   W_c(\textbf{r})&= -\frac{1}{\gamma}\int d^{3}\mathbf{q} \frac{1}{q^2}\frac{1}{b+aq^2+q^4} e^{i\mathbf{q}\mathbf{r}}\nonumber \\
    &= -\frac{2\pi}{\gamma} \int\limits_0^\infty q^2  {\frac{dq}{q^2}\frac{1}{b+aq^2+q^4}} \int\limits_0^\pi d\theta \sin{\theta} e^{iqr\cos{\theta}} \nonumber \\  
    &=-\frac{4\pi}{\gamma r}\int\limits_0^\infty dq \frac{\sin(qr)}{q(b+aq^2+q^4)}\equiv -\frac{4\pi}{\gamma r} I_0(r), \label{V_corr_sup_2}
\end{align}
where 
\begin{equation}
    \gamma = \frac{ ^2}{4m^2\omega_{p}^2},\ \ 
    b = \frac{1}{\gamma} \left( 1+\frac{1}{\epsilon_{\infty}-1} \right),\ \ 
    a = \frac{1}{\gamma}  \frac{\alpha}{q_{TF}^2}.
\end{equation}
Obviously, $I_0(r)$, as defined by Eq.~(\ref{V_corr_sup_2}), is zero for $r=0$. As the next step we differentiate $I_0(r)$ with respect to $r$ taken as parameter and make use of the 
residue theorem by properly closing contours in the complex-plane: 
\begin{align}
    I_{0}^\prime(r)&\equiv\frac{d}{dr} I_0(r)=\int\limits_{-\infty}^\infty dq \frac{\cos(qr)/2}{(b+aq^2+q^4)} \nonumber\\ 
    &=  \mathrm{Re}\left\{2\pi i \sum\limits_{\textrm{res}} \frac{e^{iqr}/2}{(b+aq^2+q^4)}\right\} \label{Eq:Appendix-I0}\\ 
    &= \mathrm{Re}\left\{\frac{\pi}{2} \Bigl(\frac{e^{-r\sqrt{\chi_1}}}{\sqrt{\chi_1}(\chi_2-\chi_1)} - \frac{e^{-r\sqrt{\chi_2}}}{\sqrt{\chi_2}(\chi_2-\chi_1)}\Bigr)\right\} \nonumber
\end{align}
with
\begin{equation}
    \chi_{1,2}=\frac{a\pm\sqrt{a^2-4b}}{2},
\end{equation}
the complex branches of the square roots should be chosen in such a way that $\mathrm{Re}[\chi_{1,2}]>0$. Integrating Eq.~(\ref{Eq:Appendix-I0}) w.r.t.~$r$ we arrive at
\begin{align}
    I_{0}(r)&=\int^r_0 I_0^\prime(u) du \\ 
    &=\mathrm{Re}\left\{\frac{\pi}{2(\chi_2-\chi_1)} \Bigl(\frac{1-e^{-r\sqrt{\chi_1}}}{\chi_1} - \frac{1-e^{-r\sqrt{\chi_2}}}{\chi_2}\Bigr)\right\}.\nonumber
\end{align}
After backward substitutions we obtain
\begin{equation} \label{V_corr_fin_sup}
    W_c(r) =- \frac{e^2}{2r}\Bigl(1 - \frac{1}{\epsilon_{\infty}} \Bigr) \mathrm{Re} \left\{ 2 - \frac{A_{+}}{A} e^{-\kappa r\sqrt{A_{-}}} + \frac{A_{-}}{A} e^{-\kappa r \sqrt{A_{+}}} \right\},
\end{equation}
where 
\begin{align} \label{A}
    A &= \sqrt{1 - \frac{2q_{TF}^2}{\alpha  \kappa^2}\frac{\epsilon_{\infty}}{\epsilon_{\infty} - 1}};\\ \kappa&=\frac{(2\alpha)^{1/2}\pi^{2/3}}{3^{1/6}}\,\rho(\textbf{r})^{1/3}; A_{\pm}=1\pm A.
\end{align}

\end{document}